# Na-Au intermetallic compounds formed under high pressure at room temperature

K. Takemura[1*] and H. Fujihisa[2]

[1] *National Institute for Materials Science (NIMS), Tsukuba, Ibaraki 305-0044, Japan*

[2] *National Institute of Advanced Industrial Science and Technology (AIST), Tsukuba, Ibaraki 305-8565, Japan*

## Abstract

High-pressure powder x-ray diffraction experiments have revealed that sodium and gold react at room temperature and form new Na-Au intermetallic compounds under high pressure. We have identified four intermetallic phases up to 60 GPa. The first phase (phase I) is the known $Na_2Au$ with the tetragonal $CuAl_2$-type structure. It changed to the second phase (phase II) at about 0.8 GPa, which has the composition $Na_3Au$ with the trigonal $Cu_3As$-type or hexagonal $Cu_3P$-type structure. Phase II further transformed to phase III at 3.6 GPa. Phase III has the same composition $Na_3Au$ with the cubic $BiF_3$-type structure. Finally, phase III changed to phase IV at around 54 GPa. Phase IV gives broad diffraction peaks, indicating large structural disorder.

## I. INTRODUCTION

Alloys of gold and alkali metals have attracted the interest of chemists early in the 20th century. Zintl *et al*. [1] obtained a black deposit from an ammonia solution of sodium by adding AuI, which they claimed to be a NaAu compound. Sommer [2] confirmed the formation of alkali metal-gold compounds, and mentioned transparent CsAu to be quite interesting. Following these studies, alkali metal-gold systems have widely been investigated. The binary alloy phase diagram for the Na-Au system shows three stoichiometric compounds, $Na_2Au$, NaAu and $NaAu_2$ [3]. The crystal structures of $Na_2Au$ and $NaAu_2$ have been clarified, while no definitive structure is given for NaAu [4]. It should be noted that these compounds are obtained from the melts of Na and Au, not by the direct reaction between the solid elements. Gold is a highly electronegative element, while alkali metals are electropositive ones. Accordingly, one may expect some ionic character for alkali metal-gold compounds. The yellow-colored transparent CsAu is a good example as mentioned above, which has later been shown to be semiconducting [5, 6]. The negative oxidation state of gold is thus a focus of intensive studies as summarized in recent review papers [7, 8].

Pressure dramatically alters the physical and chemical properties of alkali metals. Firstly, the electron density of alkali metals rapidly increases with pressure, as they are highly compressible. Secondly, the electronic structures of light (Li, Na) and heavy (K, Rb, Cs) alkali metals largely change under high pressure induced by the *s-p*, *s-d*, or *p-d* electronic transitions [9, 10]. These two changes facilitate the reactivity of alkali metals under high pressure [11]. In 1996, Atou *et al*. [12] reported the formation of $K_3Ag$ and $K_2Ag$ compounds under high pressure at room temperature by compressing the constituent elements in a diamond-anvil cell (DAC). By using the similar techniques, they further synthesized KAg$_2$ [13], and K-Ni compounds at high pressure and high temperature [14].

In the present article, we report the synthesis of Na-Au intermetallic compounds under high pressure at room temperature. The initial motivation of the present study was different. We were interested in investigating the hydrostaticity of a sodium pressure-transmitting medium by taking powder x-ray diffraction patterns of gold under pressure compressed with sodium. Any deviation from hydrostatic conditions should be detected in a systematic variation of the lattice parameters determined from different *hkl* indices, as was shown for the case of a helium pressure-transmitting medium [15]. As soon as we started the experiments, however, we found that gold reacts with sodium on application of small pressure. We thus changed the focus of the study to the Na-Au system under pressure. We identified four phases at high pressures, among which three phases are new for the Na-Au system. This finding suggests that a rich variety of phase diagrams should exist for alkali metal-gold systems under high pressure. Although we could not study the hydrostaticity of sodium by using gold itself, the cubic phase III of Na-Au gave alternative information on the hydrostaticity of sodium.

**II. EXPERIMENTAL**

We have compressed a mixture of Na and Au in a DAC and taken powder x-ray diffraction patterns at high pressures at room temperature. We used diamond anvils with 0.8 mm or 0.3 mm culet size depending on the pressure range. A fine powder of Au with average particle size of 0.7 μm (Ishifuku Metal Co. Ltd., 99.9% purity) was put in a hole in a Re gasket together with several ruby micro crystals for pressure measurement. Sodium metal (Johnson Matthey, 99.95% purity) was loaded in the gasket hole under argon atmosphere in a glove box. The volume ratio of gold to sodium was very small (1-2%). After closing the DAC, sodium in the gasket hole remained highly shiny, indicating no oxidation occurred for sodium during the sample preparation. This was also confirmed in the powder x-ray diffraction patterns, which contained no oxide peaks. Pressures were determined based on the ruby pressure scale for the He-pressure medium [16]. This could be rationalized, since sodium gives quasi-hydrostatic conditions as good as the He-pressure medium [17]. Pressures calculated from the lattice parameter of sodium based on its equation of state [18] gave consistent results with the ruby pressures. Powder x-ray diffraction experiments were done on the beamline BL-13A and AR-NE1A of the Photon Factory, KEK. Monochromatic x-ray beams with wavelength of 0.4264 Å (BL-13A) or 0.4104 Å (AR-NE1A) were collimated to 30~50 μm beam. The beam size was much smaller than the gasket hole, and hence we could choose the best position where the Na-Au intermetallic phases formed. Diffracted x-rays were recorded on an imaging plate and analyzed with the pattern integration software PIP [19]. All the experiments were done at room temperature.

Further structure analysis was done by the software package Materials Studio® from Accelrys, Inc. The cell and atomic positional parameters were first refined by the Rietveld method. The March-Dollase preferred orientation function was used for the fitting. Then, the atomic positions were optimized by the density-functional theory (DFT) software CASTEP [20] by fixing the cell parameters. We employed the GGA-PBEsol exchange-correlation functionals [21] and used an ultrasoft pseudopotential [22].

**III. RESULTS**

Three experimental runs were done up to a maximum pressure of 59.6 GPa. Diffraction data were collected on increasing and decreasing pressure cycles. The volume ratio of Au to Na was quite small (1-2%), and hence the system was always under Na-rich conditions. As soon as we started the experiments, we recognized that Au reacts with Na. The diffraction pattern taken at 0.72 GPa already showed the formation of Na-Au compounds. The minimum pressure required for the reaction was not determined in the present experiments, since some pressure was necessary to seal the gasket of the DAC. Otherwise sodium easily reacts with oxygen and water in the air. Depending on the pressure range we identified four phases, which we call phase I, II, III and IV in the order of increasing pressure. Notice that the composition of each phase is not necessarily the same. This is because, in the present experiments, Na was both a reactant and a pressure-transmitting medium surrounding the sample, thereby Na was free to move in or out of the crystal lattice.

Powder diffraction patterns taken at pressures below 3.5 GPa were complicated. Spotty diffraction peaks of Na were always present but easily distinguished from others. The rest of the diffraction peaks was explained as a mixture of phase I and II or phase II and III, except two peaks at $d = \sim 4.30$ Å and $d = \sim 2.49$ Å, which could not be indexed to any phases. We also observed faint diffraction lines of gold at low pressures. These facts mean that the reaction between gold and sodium was not complete at low pressures. No attempts were made to study the time-dependence of the reaction. At pressures higher than 3.5 GPa, we obtained clear patterns of phase III without any phase mixture. By decreasing pressure from phase III, pure diffraction patterns of single phase II and I were obtained. The diffraction rings of phases I, II, and III were sharp and smooth, probably due to the small particle size of the initial gold powder. By contrast, phase IV, existing above 54 GPa, gave broad diffraction peaks as shown later.

Figure 1(a) shows the x-ray diffraction pattern of phase I taken at 0.83 GPa on decreasing pressure. The pattern is successfully indexed by a tetragonal unit cell with cell parameters $a = 7.2817(1)$ and $c = 5.4784(1)$ Å ($c/a = 0.7524$). These parameters well correspond to those for $Na_2Au$ reported at atmospheric pressure [$a = 7.415(5)$ and $c = 5.522(5)$ Å, $c/a = 0.7447$] [23], and the pattern is well explained by the same structure. It is the $CuAl_2$-type structure with the space group $I4/mcm$ (No. 140) and the Pearson's symbol $tI12$. The unit cell contains four formula units of $Na_2Au$ ($Z = 4$). The Rietveld refinement gave reliability factors $R_{wp} = 2.04\%$, $R_{wp}$ (without background) $= 6.19\%$, and $R_p = 1.31\%$. The crystal structure is shown in Fig. 1(b). The $CuAl_2$-type structure is one of the popular structures for intermetallic compounds. The structural data and atomic positional parameters are listed in Tables I and II.

Figure 2(a) shows the diffraction pattern of phase II at 1.80 GPa taken on increasing pressure (in this particular experimental run, we obtained single phase II on increasing pressure). The pattern is more complicated than phase I. By excluding the sodium lines, the pattern was successfully indexed by a hexagonal (or trigonal) lattice with cell parameters $a = 8.7488(2)$ and $c = 9.0770(3)$ Å, and $c/a = 1.0375$. The number of atoms contained in the unit cell was determined as follows. In the compound formation induced by pressure, the volume of formed compound is smaller than, but not very far from those of the constituent elements. For example, if the unit cell of phase II contains $Z$ formula units of $Na_xAu_y$, the unit cell volume $V$ should be close to but smaller than $Z(xV_{Na} + yV_{Au})$, where $V_{Na}$ and $V_{Au}$ are the atomic volume of Na and Au at the given pressure, respectively. Among possible combinations of $Z$, $x$, and $y$, we found that $Z = 6$, $x = 3$, and $y = 1$ gives the most reasonable volume.

After several trials, two space groups $P\text{-}3c1$ (No. 165) and $P6_3cm$ (No. 185) were found to equally well explain the observed pattern. The reliability factors obtained in the Rietveld analysis were $R_{wp} = 2.90\%$, $R_{wp}$ (without background) $= 12.64\%$, and $R_p = 1.61\%$ for the former structure, and $R_{wp} = 2.85\%$, $R_{wp}$ (without background) $= 12.42\%$, and $R_p = 1.59\%$ for the latter one. The Pearson's symbol is $hP24$ for both cases. The atomic positional parameters were further optimized by DFT so as to minimize the total energy. The reliability factors for the optimized structure are, for example, $R_{wp} = 3.40\%$, $R_{wp}$ (without background) $= 14.82\%$, and $R_p = 1.82\%$ for the former ($P\text{-}3c1$). The maximum displacement of atoms from the Rietveld result was 0.46 Å. Although the reliability factors are slightly larger than the case of the Rietveld analysis, we adopt the parameters obtained in DFT, since they are energetically more favorable and the observed diffraction intensity contains comparable uncertainty (~ 5%). The final atomic positional parameters are shown in Table II.

The former structure with space group $P\text{-}3c1$ is known as the $Cu_3As$-type structure [Fig. 2(b)] [4]. The gold atoms are coordinated by 11 sodium atoms; three at distances of 2.93-3.00 Å separated approximately by 120° and situated roughly in the $ab$-plane, two at a distance of 3.02 Å nearly along the $c$-axis, and six at distances of 3.04-3.69 Å located at the corners of a distorted trigonal prism. The latter structure with space group $P6_3cm$ is known as the $Cu_3P$-type structures [Fig. 2(c)] [24]. The local atomic arrangement is similar to the $Cu_3As$-type structure, but the unit cell has hexagonal symmetry compared with the trigonal one of the $Cu_3As$-type structure. While the space group has higher symmetry, sodium atoms take four different sites with six independent positional parameters compared with three sites with four parameters in the case of Cu$_3$As-type (see Table II). It should be noted that the Cu$_3$As- and Cu$_3$P-type structures are two variants of the Li$_3$P-type structure ($P6_3/mmc$, No. 194) [25]. The trigonal prism is undistorted in the Li$_3$P-type structure, while it is distorted in a different way in the Cu$_3$As- and Cu$_3$P-type structures. Since the reliability factors are comparable for both structures, we cannot discriminate between them. Our total energy calculations also yield very small energy difference (~0.002 eV) between the two structures. We have further performed a molecular-dynamics simulation by the DFT with the isothermal-fixed volume-NVT ensemble at 300 K starting from the $P$-3$c$1 model. The simulation shows that the amplitude of the atomic vibration was approximately 0.6 Å for each atom and was comparable to the maximum displacement between the $P$-3$c$1 and $P6_3cm$ model. It is therefore difficult, at least at room temperature, to decide which structure is more plausible for phase II. This situation is similar to the case of high-pressure phase IV of Na$_3$N [26].

Figure 3(a) shows the diffraction pattern of phase III at 3.69 GPa taken on increasing pressure. The pattern is easily indexed with a cubic unit cell with $a$ = 6.9948(5) Å. By using the same criteria as used for phase II, we found that the unit cell contains four formula units of Na$_3$Au ($Z$ = 4). The structure is the BiF$_3$-type with the space group $Fm$-3$m$ (No. 225) and the Pearson's symbol $cF$16. The reliability factors were $R_{wp}$ = 2.58%, $R_{wp}$ (without background) = 8.36%, and $R_p$ = 1.49%. The structural data are summarized in Tables I and II. The crystal structure is shown in Fig. 3(b). The unit cell is made up of eight bcc sub-lattices, of which one quarter of the positions is periodically occupied by gold.

Phase III is stable over a wide pressure range up to about 52 GPa (Fig. 4). The diffraction pattern taken at higher pressures suddenly broadened, indicating a structural change. If we look at the diffraction pattern at 59.6 GPa, the first peak is located at ~6.8°, obviously shifted to lower scattering angle compared with the first peak at ~7.2° of the pattern at 51.7 GPa. Since diffraction peaks without phase transition usually shift to higher scattering angles under compression, the pattern at 59.6 GPa cannot be explained by a simple broadening of the pattern of phase III. We thus conclude that phase III undergoes a phase transition to phase IV near 54 GPa, as the diffraction pattern taken at 56.6 GPa in another experimental run was of phase IV. The broad diffraction pattern indicates large structural disorder for phase IV, but at present we have no structural model, nor explanation for the peak broadening. The composition of phase IV is unknown.

On decreasing pressure, phase IV was observed down to 18 GPa. When we released pressure down to 2.3 GPa, the sharp powder pattern of phase III appeared. Phase III successively transformed to phase II and I on further decrease of pressure (not shown in Fig. 4). Thus all the phase transitions, including the compositional change at the phase II ($Na_3Au$) to I ($Na_2Au$), were reversible. In order to study the stability of phase I, we completely released the pressure of the DAC in a glove box, closed the cell again and took diffraction pattern. Notice that we had to apply small pressure to seal the gasket. The diffraction pattern taken at 0.26 GPa after this procedure was of phase I. Phase I, once formed under pressure, seems to be stable at atmospheric pressure without decomposing into the elements. This is consistent with the phase diagram for the Na-Au system at atmospheric pressure.

Figure 5 summarizes the molar volumes of the Na-Au system together with those for Na [18] and Au [15] under pressure. The data taken on increasing and decreasing pressure agree well and are reproducible in different experimental runs. The data for phase I connect smoothly to the data obtained at atmospheric pressure [23].  The molar volume of phase I ($Na_2Au$) is smaller than that of the constituent elements, 2Na+Au, by 21% at atmospheric pressure (red arrow in Fig. 5).  Similarly, the molar volume of phase II ($Na_3Au$) is smaller than that of 3Na+Au by 15% at 0.5 GPa (blue arrow in Fig. 5).  If we compare the molar volume of phase II with that of a mixture of phase I and Na, the volume difference is about 4%.  The change from phase II to III is a structural phase transition without accompanying compositional change.  The volume decreases by about 8% at the transition.  In Fig. 5, we also show the reduced molar volume $V_m/n$, where $V_m$ is the molar volume and $n$ is the number of atoms in the formula unit for each phase (for example, $n = 3$ for phase I, $Na_2Au$) [27].  It is well known that the reduced molar volumes of alloys, solid solutions and intermetallic compounds are, in many cases, proportional to the composition ratio (the Vegard's law).  In order to see how the Na-Au system obeys the law, we show four isobaric sections of $V_m/n$ ($P$) at 1, 3, 6, and 52 GPa in Fig. 6, which are plotted as a function of concentration $x$ of Au.  All the phases I, II, and III have smaller volumes than those expected from the Vegard's law.  This implies that the bonding is stronger in Na-Au than in simple solid solutions.

The experimental pressure-volume data for phases II and III are fitted to the equation of state formulated by Vinet *et al*. [28].  The following values are obtained for the volume at atmospheric pressure ($V_0$), bulk modulus ($B_0$) and its pressure derivatives ($B_0'$): phase II [$V_0 =$ 65.7(3) $cm^3$/mol, $B_0 = 16.2(1)$ GPa, $B_0' = 4.9$ (Ref. 29)] and phase III [$V_0 = 59.1(5)$ $cm^3$/mol, $B_0 = 18.8(13)$ GPa, $B_0' = 4.9(1)$].  For comparison, sodium has a bulk modulus of 6.31 GPa [18], and gold has a bulk modulus of 167 GPa [15].  The fit was unsuccessful for phase I, because of the small pressure range.

## IV. DISCUSSION

An important finding of the present work is the formation of Na-Au compounds at room temperature under high pressure. The reaction pressure seems to be very low, but to our knowledge, no previous reports exist for the solid-solid reaction of sodium and gold. We used a fine powder of gold, which could have played a role in the compound formation. It would be interesting to study whether bulk gold also reacts with sodium at high pressure at room temperature. Potassium and silver form $K_3Ag$ with the $BiF_3$-type structure at room temperature and 6.4 GPa [12]. The situation is similar to the present case of $Na_3Au$. While $K_3Ag$ decomposes into the elements on decreasing pressure, $Na_3Au$ does not fully decompose but changes to $Na_2Au$, which is recovered at atmospheric pressure. If we compare the phase diagrams for alkali metal-gold systems, Na-Au and K-Au are similar to each other. It is very likely that the K-Au system may also form compounds and show phase changes under pressure similar to the Na-Au system. The phase diagram for Li-Au system is more complicated. We mention that $Li_3Au$ with the $BiF_3$-type structure already exists at atmospheric pressure. The phase diagrams for Rb-Au and Cs-Au systems are rather different from those for light alkali metals-gold systems, yet it should be interesting to explore the possibility of compounds formation in these systems under high pressure.

We initially expected that gold takes negatively charged ionic state in the Na-Au compounds. However, experiment showed no evidence of such a state. All the phases had shiny and metallic luster under a microscope. Our band structure calculations confirmed that phases I, II, and III should be metallic. The electronic properties of alkali-metal-gold compounds have been calculated for the 1:1 compounds assuming the CsCl-type structure [30]. The band structure calculations showed that CsAu and RbAu are semiconductors, whereas KAu, NaAu, and LiAu are metals. This is reasonable, since lighter alkali metals are less electropositive, and hence the compounds with gold should be less ionic. Similar compounds $K_2Ag$ and $K_3Ag$ are also shown to be metallic by band structure calculation [31].

We have identified two high-pressure phases II and III with the composition $Na_3Au$, which have the $Cu_3As$- or $Cu_3P$-type and $BiF_3$-type structures, respectively. These structure types are found in a number of ionic and intermetallic $AX_3$ compounds. Crystal structures similar to phase II are reported for $Cu_3As$ [32], $Na_3As$ [25, 32, 33], $Cu_3P$ [24, 34], $LaF_3$ [35], $HoD_3$ [36], $YD_3$ [37], and the high-pressure phase IV of $Na_3N$ [26]. The $BiF_3$-type structure is reported for $Li_3Bi$, $LaH_3$, $Fe_3Al$, and $Li_3Au$ at atmospheric pressure [4], and for high-pressure phases $K_3Ag$ [12], $\gamma$-$Li_3N$ [38] and phase V of $Na_3N$ [26]. The IV-V phase transition in $Na_3N$ is quite similar to the II-III phase transition in $Na_3Au$. Vajenine *et al.* [26] argue that the coordination number of N in $Na_3N$ increases from 11 in phase IV to 14 in phase V. Since the phase V of $Na_3N$ is stable up to the highest pressure investigated (36 GPa), and $\gamma$-$Li_3N$ is stable up to 200 GPa, the $BiF_3$-type structure with a high coordination number seems to be an ultimate structure for ionic $AX_3$ compounds at high pressures [39]. In this context, the transition from the $BiF_3$-type structure (phase III) to phase IV in $Na_3Au$ suggests a stability limit for this type of structure at least for intermetallic compounds under pressure.

Finally, we mention the hydrostaticity of sodium. As seen in Fig. 4, the diffraction peaks of phase III remain sharp up to 52 GPa just before the transition to phase IV, showing that the sodium pressure medium offers good quasi-hydrostatic conditions. The lattice parameters of phase III determined from different *hkl* reflections agree well within 0.04% in our experiments. These observations are consistent with the results by Hanfland *et al.* [17], which show sharp diffraction peaks of Ta in a sodium pressure medium up to 69 GPa with a maximum deviation of *d*-values by 0.02%. Sodium can be used as a good quasi-hydrostatic pressure-medium, unless it reacts with the sample as in the present case.

In summary, we have observed that sodium and gold react at room temperature at high pressures. Four Na-Au intermetallic phases have been identified, of which three phases (phase I, II, and III) have typical crystal structures for $AX_2$ and $AX_3$ compounds. Phase III undergoes a transition to phase IV at higher pressures. The structure is unknown but seems to be highly disordered, possibly being a new structure type for the $AX_3$ compounds under pressure. Alkali metal-gold systems should have a rich variety of phase diagrams under high pressure.

**Acknowledgements**

We would like to thank Yu. Grin and W. A. Crichton for valuable discussions. The experiments at the Photon Factory were done under the proposal number 2008G534.

**References**

[*] Electronic address: takemura.kenichi@nims.go.jp

[1] E. Zintl, J. Goubeau, and W. Dullenkopf, Z. Phys. Chem. **154A**, 1 (1931).

[2] A. Sommer, Nature **152**, 215 (1943).

[3] T. B. Massalski, H. Okamoto, P. R. Subramanian, and L. Kacprzak, *Binary Alloy Phase Diagrams*, second ed. (ASM International, Materials Park, OH, 1990).

[4] P. Villars and L. D. Calvert, *Pearson's Handbook of Crystallographic Data for Intermetallic Phases*, second ed. (ASM International, Materials Park, OH, 1991).

[5] W. E. Spicer, A. H. Sommer, and J. G. White, Phys. Rev. **115**, 57 (1959).

[6] F. Wooten and G. A. Condas, Phys. Rev. **131**, 657 (1963).

[7] J. E. Ellis, Inorg. Chem. **45**, 3167 (2006).

[8] M. Jansen, Chem. Soc. Rev. **37**, 1826 (2008).

[9] N. E. Christensen and D. L. Novikov, Solid State Commun. **119**, 477 (2001).

[10] M. Ross and A. K. McMahan, Phys. Rev. B **26**, 4088 (1982).

[11] J. V. Badding, L. J. Parker, and D. C. Nesting, J. Solid State Chem. **117**, 229 (1995).

[12] T. Atou, M. Hasegawa, L. J. Parker, and J. V. Badding, J. Am. Chem. Soc. **118**, 12104 (1996).

[13] M. Hasegawa, T. Atou, and J. V. Badding, J. Solid State Chem. **130**, 311 (1997).

[14] L. J. Parker, T. Atou, and J. V. Badding, Science **273**, 95 (1996).

[15] K. Takemura and A. Dewaele, Phys. Rev. B **78**, 104119 (2008).

[16] C.-S. Zha, H.-k. Mao, and R. J. Hemley, Proc. Nat. Acad. Sci. U.S.A. **97**, 13494 (2000).

[17] M. Hanfland, K. Syassen, and J. Kohler, J. Appl. Phys. **91**, 4143 (2002).

[18] M. Hanfland, I. Loa, and K. Syassen, Phys. Rev. B **65**, 184109 (2002).

[19] H. Fujihisa (unpublished).

[20] S. J. Clark, M. D. Segall, C. J. Pickard, P. J. Hasnip, M. I. J. Probert, K. Refson, and M. C. Payne, Z. Kristallogr. **220**, 567 (2005).

[21] J. P. Perdew, A. Ruzsinszky, G. I. Csonka, O. A. Vydrov, G. E. Scuseria, L. A. Constantin, X. Zhou, and K. Burke, Phys. Rev. Lett. **100**, 136406 (2008).

[22] D. Vanderbilt, Phys. Rev. B **41**, 7892 (1990).

[23] E. E. Havinga, H. Damsma, and P. Hokkeling, J. Less-Common Metals **27**, 169 (1972).

[24] O. Olofsson, Acta Chem. Scand. **26**, 2777 (1972).

[25] G. Brauer and E. Zintl, Z. Physik. Chem. [Abt. B] **37**, 323 (1937).

[26] G. V. Vajenine, X. Wang, I. Efthimiopoulos, S. Karmakar, K. Syassen, and M. Hanfland, Phys. Rev. B **79**, 224107 (2009).

[27] G. Kienast and J. Verma, Z. Anorg. Allgem. Chem. **310**, 143 (1961).

[28] P. Vinet, J. Ferrante, J. H. Rose, and J. R. Smith, J. Geophys. Res. **92**, 9319 (1987).

[29] The value of $B_0'$ for phase II was fixed to the same one as for phase III in order to get good convergence of the fit.

[30] C. Koenig, N. E. Christensen, and J. Kollar, Phys. Rev. B **29**, 6481 (1984).

[31] J. S. Tse, G. Frapper, A. Ker, R. Rousseau, and D. D. Klug, Phys. Rev. Lett. **82**, 4472 (1999).

[32] M. Mansmann, Z. Kristallogr. **122**, 399 (1965).

[33] P. Hafner and K.-J. Range, J. Alloys Comp. **216**, 7 (1994).

[34] B. Steenberg, Arkiv Kemi **12**A, 1 (1938).

[35] M. Mansmann, Z. Kristallogr. **122**, 375 (1965).

[36] M. Mansmann and W. E. Wallace, J. Phys. **25**, 4549 (1964).

[37] T. J. Udovic, Q. Huang, and J. J. Rush, J. Phys. Chem. Solids **57**, 423 (1996).

[38] A. Lazicki, B. Maddox, W. J. Evans, C.-S Yoo, A. K. McMahan, W. E. Pickett, R. T. Scalettar, M. Y. Hu, and P. Chow, Phys. Rev. Lett. **95**, 165503 (2005).

[39] W. A. Crichton, P. Bouvier, B. Winkler, and A. Grzechnik, Dalton Trans. **39**, 4302 (2010).

(a)

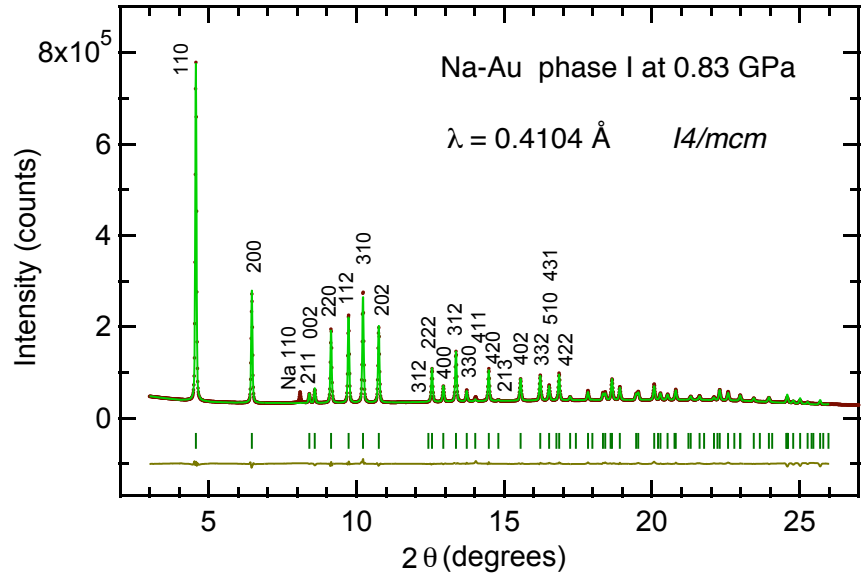

(b)

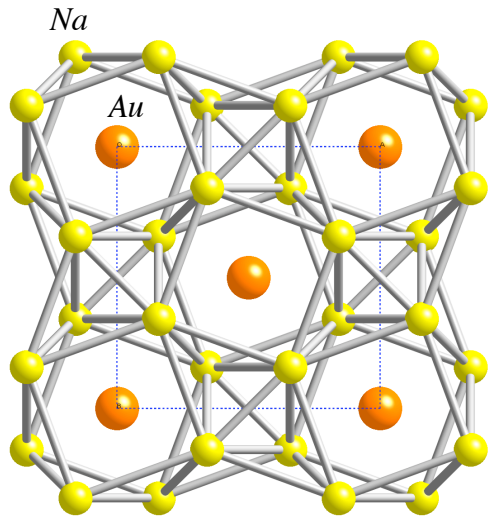

FIG. 1: (Color online) Phase I (Na$_2$Au) of Na-Au at 0.83 GPa. (a) Powder x-ray diffraction pattern. The brown points show the observed data, and the green line shows the Rietveld refinement. The green ticks below the pattern show the calculated peak positions. The difference between the observed and calculated intensity is shown by light green line at the bottom. The *hkl* indices are shown for the major reflections. The 110 reflection of Na is also observed. (b) The CuAl$_2$-type structure projected onto the *a-b* plane.

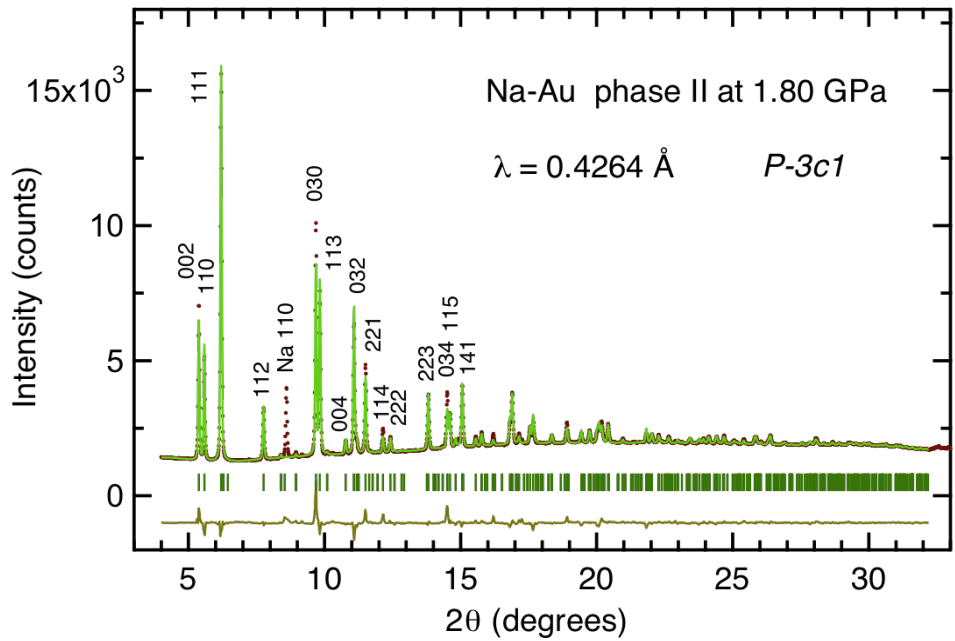

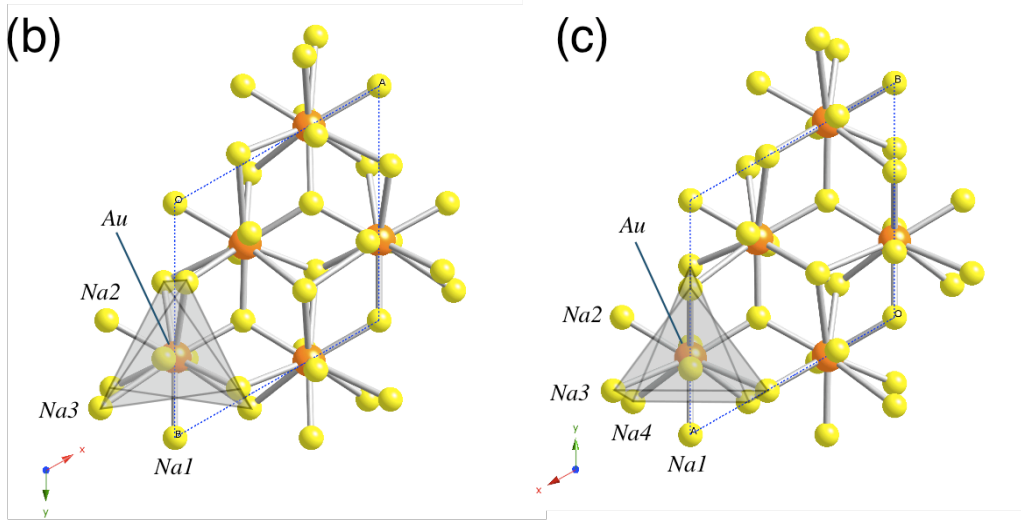

FIG. 2: (Color online) Phase II ($Na_3Au$) of Na-Au at 1.80 GPa. (a) Powder x-ray diffraction pattern and the Rietveld refinement as the $Cu_3As$-type structure (see the caption to Fig. 1 for the legend). (b) The $Cu_3As$-type structure projected onto the *a-b* plane. The hatched shape at the lower left corner shows the distorted trigonal prism. (c) The $Cu_3P$-type structure.

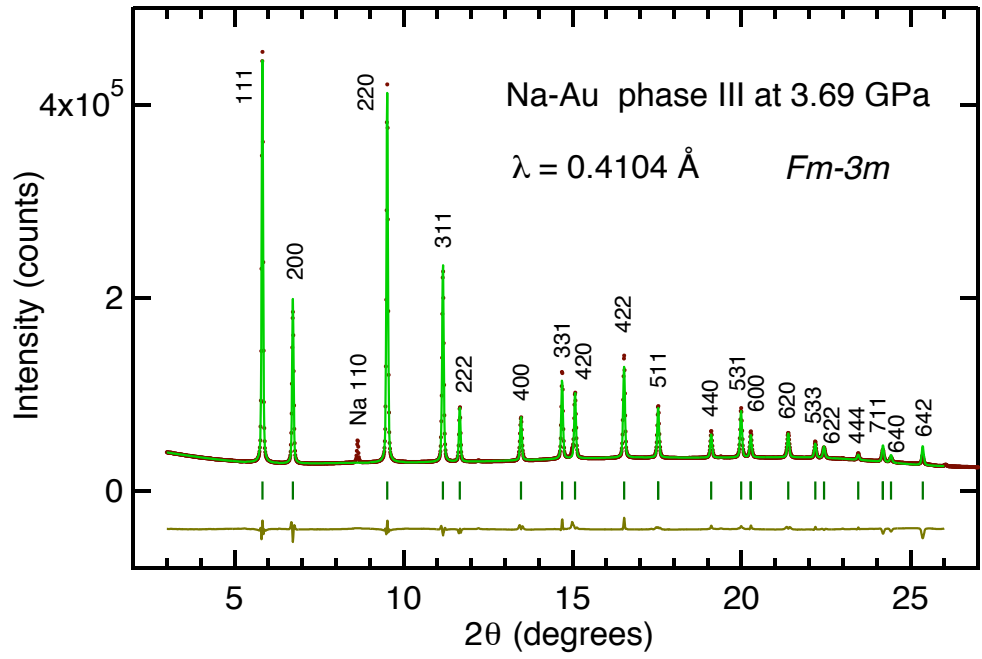

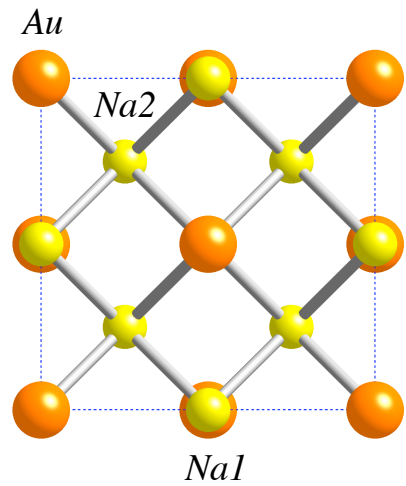

FIG. 3: (Color online) Phase III ($Na_3Au$) of Na-Au at 3.69 GPa. (a) Powder x-ray diffraction pattern (see the caption to Fig. 1 for the legend). (b) The $BiF_3$-type structure projected onto the $a$-$b$ plane.

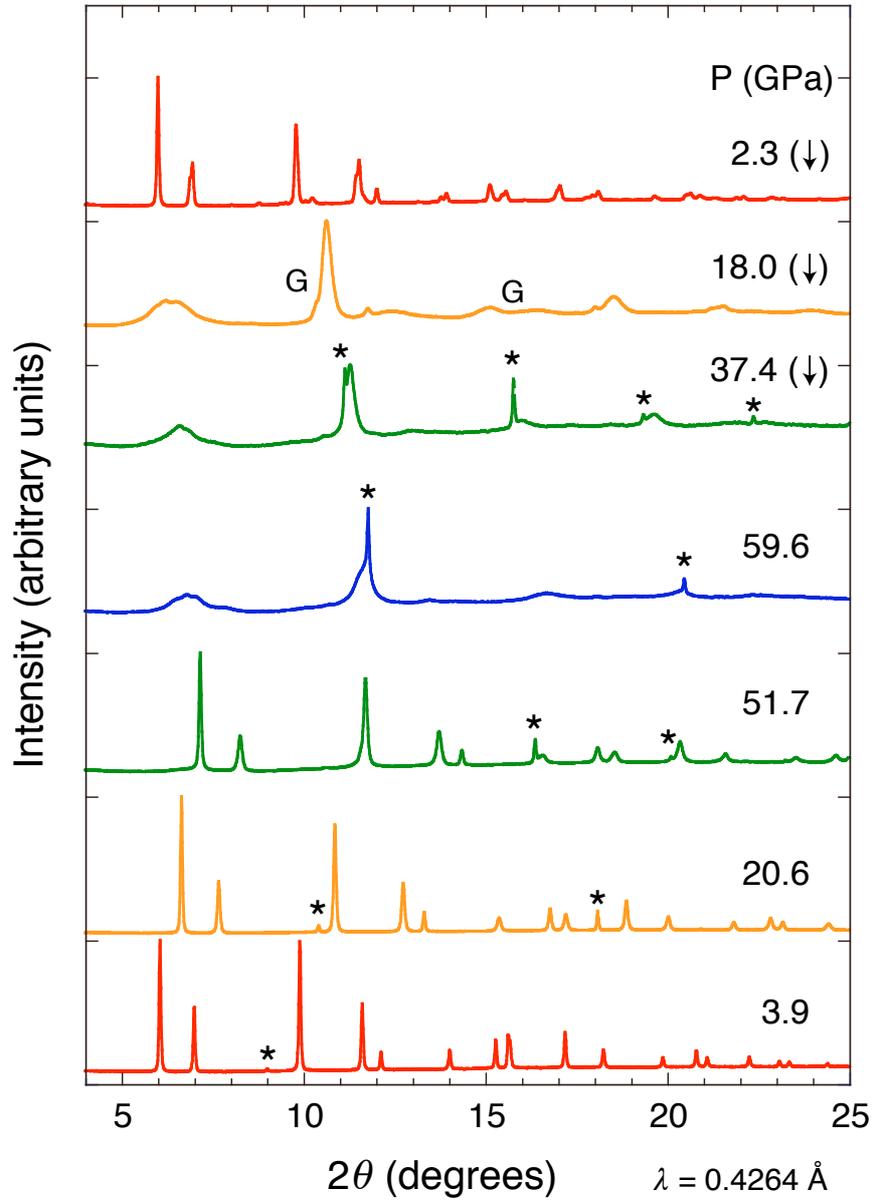

FIG. 4: (Color online) Powder x-ray diffraction patterns of phases III and IV. The pressure values with downward arrows indicate the patterns taken in the course of decreasing pressure. Phase III persists up to about 52 GPa, then transforms to phase IV with broad diffraction peaks as shown in the pattern at 59.6 GPa. On decreasing pressure, phase IV was observed to at least 18 GPa. It transformed back to phase III at 2.3 GPa. The sharp peaks marked by asterisk are the diffraction peaks of Na. $G$ indicates the diffraction peaks of the Re gasket. Horizontal bars on the vertical axis indicate the zero level for each pattern.

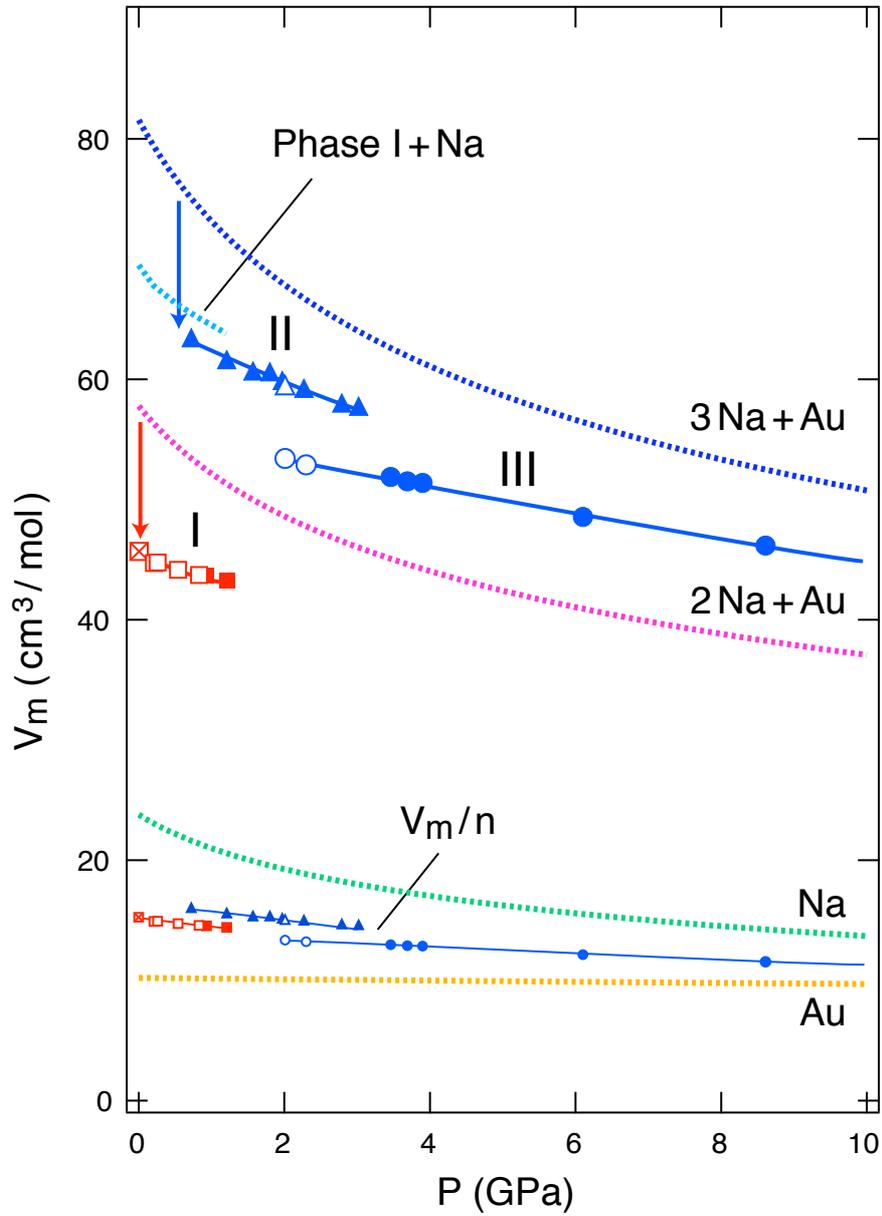

FIG. 5: (Color online) Molar volume of phases I, II, and III of Na-Au intermetallic compound as a function of pressure. Closed symbols show the data taken on increasing pressure, and open symbols on decreasing pressure. The atmospheric data shown by crossed square is from Ref. 23. The molar volumes of Na (Ref. 18), Au (Ref. 15) and their mixtures are shown by dotted curves for comparison. Smaller symbols show the molar volumes divided by the number of atoms in a formula unit for respective phases (see text).

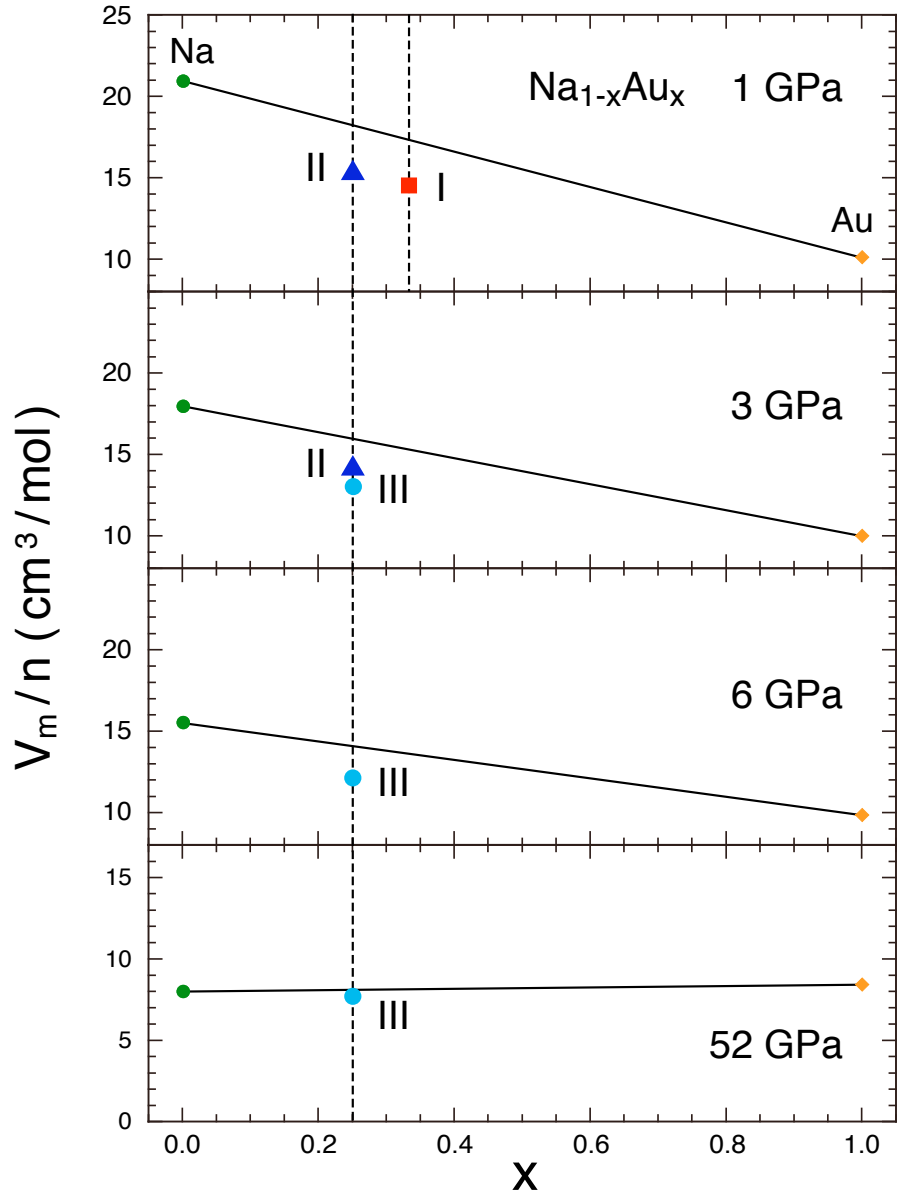

FIG. 6: (Color online) Reduced molar volume of $Na_{1-x}Au_x$ as a function of concentration $x$ of Au at high pressures. Phases I, II, and III have smaller volumes than those expected from the Vegard's law, which are shown by the straight lines connecting Na and Au.

# Tables

TABLE I: Structural data for phase I, II, and III of Na-Au compounds under high pressure. [a]Reference 23. *Data taken on decreasing pressure.

| Phase | Formula | Structure type | Space group, Z | P (GPa) | a (Å) | c (Å) | V (Å$^3$) |
|---|---|---|---|---|---|---|---|
| I | Na$_2$Au | CuAl$_2$ | $I4/mcm$, 4 | 0.0 [a] | 7.415(5) | 5.522(5) | 303.6 (7) |
|  |  |  |  | 0.83* | 7.2817(1) | 5.4784(1) | 290.49(1) |
| II | Na$_3$Au | Cu$_3$As or Cu$_3$P | $P\text{-}3c1$ or $P6_3cm$, 6 | 1.80 | 8.7488(2) | 9.0770(3) | 601.69(5) |
| III | Na$_3$Au | BiF$_3$ | $Fm\text{-}3m$, 4 | 3.69 | 6.9948(5) |  | 342.24(7) |
|  |  |  |  | 8.61 | 6.744(1) |  | 306.7(1) |
|  |  |  |  | 15.3 | 6.510(2) |  | 275.9(2) |
|  |  |  |  | 20.6 | 6.371(2) |  | 258.6(2) |
|  |  |  |  | 29.3 | 6.205(2) |  | 239.0(2) |
|  |  |  |  | 39.0 | 6.056(1) |  | 222.1(1) |
|  |  |  |  | 51.7 | 5.915(3) |  | 206.9(3) |

TABLE II: Atomic positional parameters refined by DFT for phase I, II, and III of Na-Au compound under high pressure. The $z$-parameter for Au of phase II as the $Cu_3P$–type structure is set to 1/4 so that the comparison with the $Cu_3As$–type becomes easy.

| Atom | Site | $x$ | $y$ | $z$ |
|------|------|-----|-----|-----|
| Phase I at 0.83 GPa, $CuAl_2$–type ($I4/mcm$) | | | | |
| Na | 8$h$ | 0.170 | 0.670 | 0 |
| Au | 4$a$ | 0 | 0 | 1/4 |
| Phase II at 1.80 GPa as $Cu_3As$–type ($P$-3$c$1) | | | | |
| Na1 | 2$a$ | 0 | 0 | 1/4 |
| Na2 | 4$d$ | 1/3 | 2/3 | 0.315 |
| Na3 | 12$g$ | 0.310 | 0.945 | 0.079 |
| Au | 6$f$ | 0.658 | 0 | 1/4 |
| Phase II at 1.80 GPa as $Cu_3P$–type ($P6_3cm$) | | | | |
| Na1 | 2$a$ | 0 | 0 | 0.180 |
| Na2 | 4$b$ | 1/3 | 2/3 | 0.296 |
| Na3 | 6$c$ | 0.378 | 0 | 0.579 |
| Na4 | 6$c$ | 0.280 | 0 | 0.921 |
| Au | 6$c$ | 0.327 | 0 | 0.250 |
| Phase III at 3.69GPa, $BiF_3$–type ($Fm$-3$m$) | | | | |
| Na1 | 4$b$ | 1/2 | 1/2 | 1/2 |
| Na2 | 8$c$ | 1/4 | 1/4 | 1/4 |
| Au | 4$a$ | 0 | 0 | 0 |